# Portuguese eyewitness accounts of the great space weather event of 1582

V.M.S. Carrasco[1,2], J.M. Vaquero[2,3]

[1] Departamento de Física, Universidad de Extremadura, 06071 Badajoz, Spain [e-mail: vmscarrasco@unex.es]

[2] Instituto Universitario de Investigación del Agua, Cambio Climático y Sostenibilidad (IACYS), Universidad de Extremadura, 06006 Badajoz, Spain

[3] Departamento de Física, Universidad de Extremadura, 06800 Mérida, Spain

**Abstract:** Newly discovered descriptions about the great aurora observed in March 1582 are presented in this work. These records were made by Portuguese observers from Lisbon. Both records described the aurora like a great fire in the northern part of the sky. It was observed during three consecutive nights, according to one of the sources. Thus, we present a discussion of these auroral records in order to complement other works that studied the aurora sighted in March 1582.

## 1. Introduction

The interest in space weather and climate has been increased because of its influence on our planet in general and on our technological society in particular. Solar activity can be shown in multiple ways (Usoskin 2017). Some of them would be the number of sunspots appearing on the solar photosphere and the formation of auroras in the atmosphere of our planet (Vaquero & Vázquez 2009). Thus, auroras can be used as a proxy for the study of past solar activity.

Aurora has been a phenomenon of general interest (not only from a scientific point of view) and, for that reason, it has been recorded for millennia. There are available catalogues including auroral records made millennia ago (for example, Fritz 1873; Yau & Stephenson 1995). Recently, Hayakawa et al. (2019a) have identified in the Assyrian astrological reports the earliest candidate recorded around 660 BCE. Furthermore, several extreme space weather events have been studied from the auroral and geomagnetic records. For example, Vázquez et al. (2016) studied auroral events from 1600 to the present finding significant long-term variations in the space-time distribution of auroras and Lefèvre et al. (2016) showed a detailed analysis of data related with historical extreme geomagnetic storms for the period 1868–2010. We highlight that auroras were even sighted lower than 20º in latitude during the Carrington event in 1859 which is considered as one of the most extreme space weather events reported (Carrington 1859; Cliver & Svalgaard 2004; Hayakawa et al. 2019b).

Another example of an extreme space weather event occurred in 1582. Hattori et al. (2019), based on auroral records in East Asia, estimated that this severe storm occurred on 8 March 1582 and was comparable with other more recent magnetic storms as, for example, those in 1909 (Hayakawa et al. 2019c) and 1989 (Allent et al. 1989). Hattori et al. (2019) also indicated that the duration of this storm could be three days. Very few records from South Europe are available for this event. In fact, the only record made in the Iberian Peninsula corresponds to one auroral record made in Madrid on 6 March 1582 (Hattori et al. 2019). We here present two auroral records made in Portugal of this severe space weather event in order to complement the space-time distribution of the study made by Hattori et al. (2019). Section 2 includes the original records and a brief discussion is given in Section 3, and conclusion in Section 4.

## 2. Auroral Records

We found two records made in Portugal that include descriptions about the aurora of March 1582. We note that dates included in these records are in Julian calendar, the current calendar at that time. The first one is preserved at the District Archive of Évora [Arquivo Distrital de Évora], Portugal. In the manuscript "Cod. Cv/1-27 d.", we can read the following text: "[Lisboa] Na era de 1582 aos seis dias do mes de março apareceo o grande foguo no ceo ha parte do norte e durou tres noites" [English translation is: "[Lisbon] In 1582, on 6 March, a great fire appeared in the sky at the north and lasted three nights"].

Other Portuguese records about this aurora can be found in a manuscript entitled "Memorial de Pero Ruiz Soares", a Portuguese chronicle by Soares (1953, p. 200). In this case, the description does not include the exact Portuguese location where this aurora was sighted. However, we can see Soares (1953) indicates in the preface that noteworthy which occurred at Lisbon are described in this documentary source. Thus, we can suppose that the record of this aurora was made in Lisbon. The transcription of the auroral records included in this documentary source is: "[…] em março logo seguinte da era de 1582 hua terçafrᵃ a noute as 8 oras Comesou no çeo na banda do norte […] toda aquella parte do çeo ardendo em chamas de fogo que paressia arder ho mesmo çeo […] e naõ auia pessoa que se acordasse uer outro tal […] sobre a meya noute ueyo ter sobre o Castelo donde botou grãdes Rayos de labaredas de fogo que metia pauor e medo e logo ao outro dia fes o mesmo e as mesmas oras mas ia não taõ grande nem tam temerario e naõ auia pessoa que naõ saisse ao Campo a uer este tam grande sinal […]". The English translation of this description is: "[…] in March 1582, at 8 p.m. on Tuesday [6th March], something started in the north band of the sky […] all that part of the sky appeared burning in fire flames; it seemed that the sky was burning […] and nobody remembered having seen something like that […] at midnight, great fire rays arose above the castle which were dreadful and fearful. The following day, it happened the same at the same hour but it was not so great and terryfing. Everybody went to the countryside to see this great sign […]"

**3. Analysis and Discussion**

Auroras can be observed occasionally at mid and low latitudes due to strong geomagnetic storms (Carrington 1859; Allent et al. 1989; Akasofu 2007; Hayakawa et al. 2018). In the Iberian Peninsula, we can highlight the systematic records of auroras observed in Lisbon in the late 18th century (Vaquero & Trigo 2005) and in Barcelona during the period 1780–1825 (Vaquero et al. 2010). Moreover, there is a catalogue published by Vaquero et al. (2003) about the auroras sighted in the Iberian Peninsula for the 18[th] century and the first half of the 19[th] century (1700–1855) and also a catalogue by Aragonès & Ordaz (2010) covering only the 18th Century. Other auroral observations made in the Iberian Peninsula were studied by Farrona et al. (2011) and Carrasco et al. (2017, 2018).

According to both documentary sources reported in this work, the aurora observed in March 1582 was sighted in Lisbon. Figure 1 depicts the European places where aurora was recorded according to Hattori et al. (2019) in addition to the records made in Lisbon reported here. It implies that the records made in Lisbon were the southernmost observation recorded in Europe about the aurora of 1582. We note that this fact does not mean the aurora was necessarily at the latitude of Lisbon since aurora seen at Lisbon could be occurring a few hundred kilometers north of Lisbon. Both descriptions reported that the event started on 6 March and could be seen during three nights (according to the first record presented) and two nights (according to the manuscript by Pero Ruiz Soares). Furthermore, this last report indicates that the start time of this event was 8 p.m. (local time) on 6 March 1582. We note that the full moon closest to the period 6–8 March 1582 was on 9 March 1582. Thus, we can consider that this aurora was sighted during a period of almost full moon. This fact gives an idea of the light of the aurora. The two records situated the aurora in the northern sky and described the aurora like a large fire increasing its size during the night. Therefore, we can deduce that the color of the aurora was reddish. Typical auroras show red color of a forbidden line of atomic oxygen at 630.0 nm at high altitudes (>200 km) and, moreover, the stable auroral red arcs can be visible at mid latitude in large geomagnetic storms (Rees and Roble 1975; Kozyra et al. 1997; Hayakawa et al. 2015). We highlight that the second auroral description presented indicates that this aurora was sighted during two nights but the brightness of the auroral light during the second night was lower than in the first one. Furthermore, according to the five criteria defined by Neuhäuser & Neuhäuser (2015) to classify the likehood of the event to be an aurora (color, auroral motion, direction, night-time observation, and repetition), the descriptions here presented fulfill all criteria. Therefore, the observation of an aurora would be "almost certain" according to Neuhäuser & Neuhäuser (2015) classification.

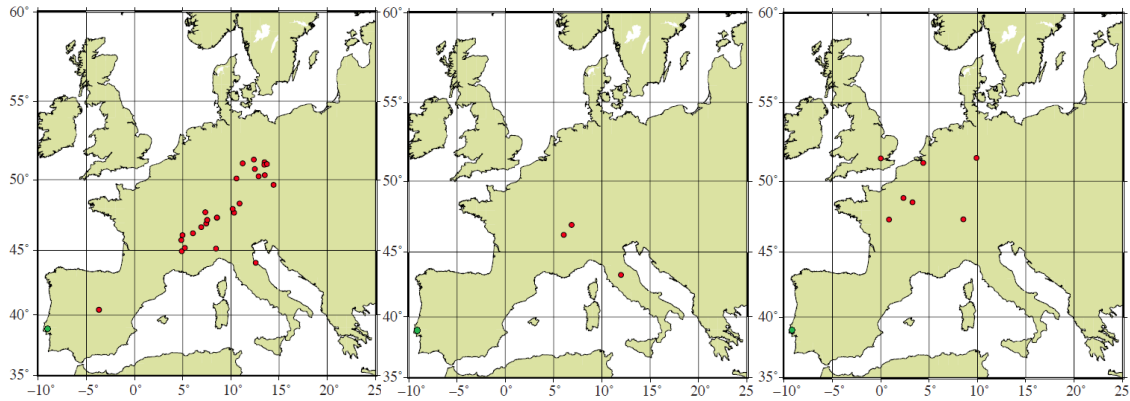

Figure 1. European locations where aurora was recorded on 6 (left panel), 7 (middle panel), and 8 (right panel) March 1582 according to Hattori et al. (2019) (red dots) and this work (green dots) (modified from Hattori et al. 2019).

## 4. Conclusions

We have found two texts containing descriptions of the great aurora sighted in March 1582 in Lisbon (Portugal). We present a discussion of these records in order to complement the work about this space weather event made by Hattori et al. (2019). Lisbon would represent the southernmost European location where this aurora was recorded according to the historical data available, together with Madrid where this aurora was also observed. The aurora was observed during three and two nights in accordance to the first and second record presented, respectively. According to both descriptions, the aurora was observed in the northern part of the sky like a fire flames. Therefore, we deduce that the color of the aurora was red. It occurred during a full moon period and this fact would indicate a large brightness of the aurora resulting from the geomagnetic storm. Thus, the two descriptions of the aurora agree with the most favorable criterion ("almost certain") regarding the aurora sightings according to Neuhäuser & Neuhäuser (2015) classification.

## Acknowledgements

This research was supported by the Economy and Infrastructure Counselling of the Junta of Extremadura through project IB16127 and grant GR18097 (co-financed by the